\documentclass[english]{article}
\usepackage{amsmath}
\usepackage{graphicx}
\usepackage[latin9]{inputenc}

\makeatletter

\providecommand{\tabularnewline}{\\}

\@ifundefined{showcaptionsetup}{}{%
 \PassOptionsToPackage{caption=false}{subfig}}
\usepackage{subfig}
\makeatother

\usepackage{babel}
\begin{document}
\title{Computationally~efficient~transfinite~patches with~fullness~control}
\author{Péter Salvi, István Kovács and Tamás Várady\\
Budapest University of Technology and Economics}
\maketitle
\begin{abstract}
Transfinite patches provide a simple and elegant solution to the problem
of representing non-four-sided continuous surfaces, which are useful
in a variety of applications, such as curve network based design.
Real-time responsiveness is essential in this context, and thus reducing
the computation cost is an important concern. The Midpoint Coons (MC)
patch presented in this paper is a fusion of two previous transfinite
schemes, combining the speed of one with the superior control mechanism
of the other. This is achieved using a new constrained parameterization
based on generalized barycentric coordinates and transfinite blending
functions.

\end{abstract}

\section{\label{sec:Introduction}Introduction}

The representation of multi-sided surfaces is a difficult problem,
with different solutions suited to different applications. Most CAD
systems provide trimmed tensor product surfaces to reproduce patches
of arbitrary sides in a mechanical model, while in computer graphics
subdivision surfaces are the de facto standard. Splitting multi-sided
regions into quadrilateral patches\textemdash preserving continuity
along splitting curves\textemdash is also a well-researched approach.

Transfinite interpolation has many advantages over the above methods:
(i)~it retains the continuity of its edge curves on the whole domain,
while exactly interpolates $G^{1}$ boundary conditions; (ii)~it
depends only on the boundary data, without the need of additional
control points or polyhedra; (iii)~smooth connections to adjacent
surfaces are easily ensured.

There are also some limitations: (i)~as the whole patch is being
defined as a blend of its boundaries, there is little control over
the center of the surface; (ii)~computation costs are somewhat higher
compared to conventional techniques; (iii)~the computation of derivatives
is complex (normally discrete approximations are used).

One of these concerns, central control, has been alleviated recently
by the addition of an extra degree of freedom (adjusting surface fullness)
to the Gregory patch~\cite{Gregory:1986}, the most popular choice
of multi-sided transfinite patch~\cite{Salvi:2016:GrafGeo}. As for
computational complexity, a practically equivalent, but much more
efficient formulation has been given for the same surface~\cite{Salvi:2014,Salvi:2015:KEPAF}.
These two modifications of the Gregory patch are, however, not compatible.
The problem lies in the parameterization of the domain, which is enhanced
in this paper in such a way as to accommodate for both adjustments,
and thus create an efficient representation capable of fullness control.

The rest of the paper is organized as follows. In Section~\ref{sec:Previous-work}
we give a short review of related research. In Section~\ref{sec:Preliminaries}
the necessary details of the above two transfinite surfaces are presented.
The new Midpoint Coons (MC) patch is introduced in Section~\ref{sec:Midpoint-Coons-patches},
and test results are shown in Section~\ref{sec:Test-results}.

\section{\label{sec:Previous-work}Previous work}

\begin{figure*}
\begin{centering}
\includegraphics[width=\textwidth]{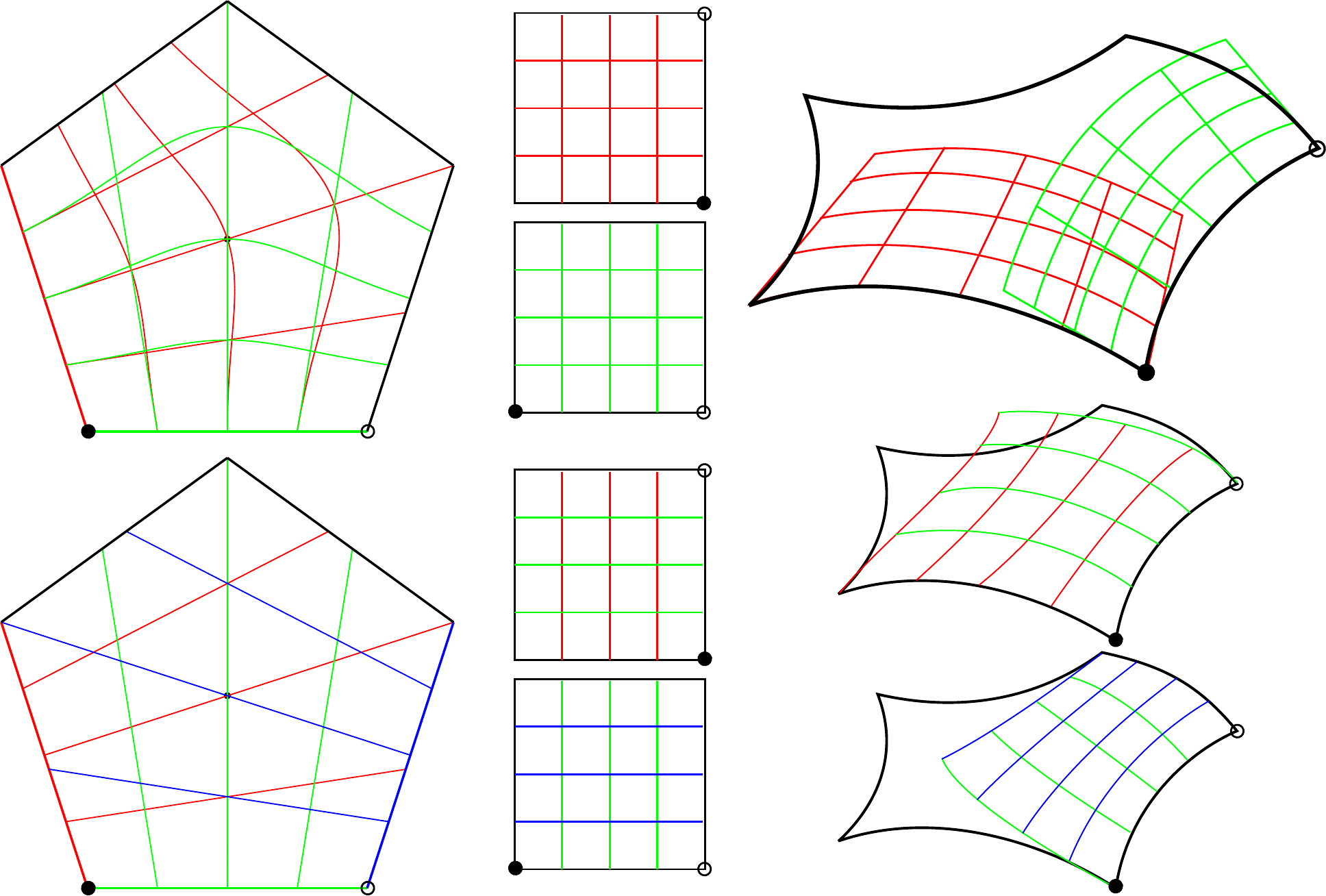}
\par\end{centering}
\caption{\label{fig:Side--and-corner-based}Side- and corner-based transfinite
surface interpolation. Side interpolants (top) are parameterized by
a side- and a distance parameter, while corner interpolants (bottom)
use only side parameters. A point in the polygonal domain (left) is
mapped to the four-sided domain of each of its ribbons (middle), which
are evaluated in 3D space (right) and then blended together, giving
a point of the transfinite patch.}
\end{figure*}
There has not been much work done on the interior control of transfinite
interpolation surfaces, maybe because these were regarded as the means
of filling multi-sided holes, not as a design tool. One exception
is an earlier work of the authors~\cite{Varady:2012}, where the patch
center can be adjusted via auxiliary surfaces. The formulation is
a modification of Kato's patch~\cite{Kato:1991}.

As briefly outlined above, this paper mainly draws on two previous
representations: the Generalized Coons (GC) and Midpoint (MP) surfaces~\cite{Salvi:2014,Salvi:2016:GrafGeo}.
The former is a multi-sided generalization of the Coons patch, which
is shown to be virtually the same as the Gregory patch~\cite{Salvi:2015:KEPAF},
while the latter introduces a central control point for fullness control.
Both of these are presented in detail in Section~\ref{sec:Preliminaries}.

Our primary contribution, the constrained parameterization in Section~\ref{subsec:Constrained-barycentric},
is based on generalized barycentric coordinates (see e.g.~Floater~\cite{Floater:2015})
and the blending function of Kato's patch~\cite{Kato:1991}.

\section{\label{sec:Preliminaries}Preliminaries}

In this section we will first look at the various constituents of
transfinite patches, then review some concrete constructions: the
Gregory patch, and its two enhancements, the Generalized Coons (GC)
and Midpoint (MP) patches.

Given a loop of 3D curves and cross-derivatives, we want to generate
a surface that interpolates these boundary conditions (in a $G^{1}$
sense, i.e., only tangent planes are reproduced). Cross-derivatives
can be defined automatically using a frame sweep (e.g.~with a rotation-minimizing
frame~\cite{Wang:2008}), or semi-automatically by first fixing normal
vectors at arbitrary points. We assume that the cross-derivative functions
are twist-compatible; otherwise rational twists can be applied~\cite{Gregory:1974}.

Transfinite surfaces come in two flavors: side-based and corner-based
schemes, see Figure~\ref{fig:Side--and-corner-based}. Side-based
schemes blend together \emph{side-interpolants} (four-sided surfaces
that interpolate one side), while corner-based schemes use \emph{corner-interpolants}
(four-sided surfaces that interpolate two adjacent sides). In practice
it is much easier to construct side-interpolants, so we will define
corner-interpolants based on these, as well.

A side-interpolant, or \emph{ribbon}, for side $i$ is defined as
\begin{equation}
R_{i}(s_{i},d_{i})=P_{i}(s_{i})+\gamma(d_{i})T_{i}(s_{i}),
\end{equation}
where $P_{i}(s_{i})$ is the $i$-th boundary curve parameterized
in $[0,1]$, $T_{i}$ is the corresponding cross-derivative, and $\gamma$
is a scaling function. (A recommended $\gamma$ function is $\gamma(d)=d/(2d+1)$,
its derivation can be found in Salvi~et~al.~\cite{Salvi:2014}) The
arguments $s_{i}$ and $d_{i}$ are the side- and distance-parameters,
with
\begin{align}
s_{i} & \in[0,1], & d_{i} & \geq0.
\end{align}

The $n$-sided patch itself is defined over a convex polygonal domain,
e.g.~a regular $n$-sided polygon in the $(u,v)$ plane. A crucial
component of a transfinite scheme is the parameter mapping from $(u,v)$
to $(s_{i},d_{i})$, i.e., from the $n$-sided polygon to each ribbon's
own parameterization. A basic constraint is that for a point on side
$i$ of the domain polygon,
\begin{align}
s_{i-1} & =1, & s_{i+1} & =0, & d_{i} & =0.
\end{align}
Also, the side parameter $s_{i}$ changes linearly from 0 to 1, and
the distance parameter $d_{i}$ grows monotonically as we go inside
the domain.

Finally, we will need suitable blending functions that interpolate
the ribbons at the boundaries, but blend them together inside the
patch.

\subsection{\label{subsec:Gregory-patch}Gregory patch}

The classic Gregory patch is a corner-based\footnote{Because of this, it is also called ``CB patch''~\cite{Salvi:2014}.}
scheme, so our first step is the creation of corner interpolants.
These can be constructed as the Boolean sum of two adjacent ribbons,
with the common part subtracted:
\begin{align}
I_{i,i-1}(u,v) & =R_{i-1}(s_{i-1},s_{i})+R_{i}(s_{i},1-s_{i-1})\nonumber \\
 & -Q_{i,i-1}(s_{i},s_{i-1}),
\end{align}
where the $Q_{i,i-1}$ correction patch is defined as
\begin{align}
Q_{i,i-1}(s_{i},s_{i-1}) & =P_{i}(0)+\gamma(1-s_{i-1})T_{i}(0)+\gamma(s_{i})T_{i-1}(1)\nonumber \\
 & +\gamma(s_{i})\gamma(1-s_{i-1})W_{i,i-1}.
\end{align}
Here $W_{i,i-1}$ is the (common) twist for the corner $(i,i-1)$.

The patch equation is simply
\begin{equation}
S_{CB}(u,v)=\sum_{i=1}^{n}I_{i,i-1}(u,v)B_{i,i-1}(u,v),
\end{equation}
where $B_{i,i-1}$ is the blending function
\begin{equation}
B_{i,i-1}(u,v)=\frac{D_{i,i-1}}{\sum_{j=1}^{n}D_{j,j-1}}=\frac{1/(d_{i}d_{i-1})^{2}}{\sum_{j=1}^{n}1/(d_{j}d_{j-1})^{2}}
\end{equation}
with $D_{i,i-1}=\prod_{k\notin\{i,i-1\}}d_{k}^{2}$.

For parameterization, radial side parameters and perpendicular distance
parameters are used. As these are not relevant to the enhancements
at hand, the reader is referred to the original paper~\cite{Gregory:1986}.

\subsection{\label{subsec:Generalized-Coons-patch}Generalized Coons patch}

This is a side-based formulation, similar in logic to the original
four-sided Coons patch:
\begin{align}
S_{GC}(u,v) & =\sum_{i=1}^{n}R_{i}(s_{i},d_{i})B_{i}(u,v)\nonumber \\
 & -\sum_{i=1}^{n}Q_{i,i-1}(s_{i},s_{i-1})B_{i,i-1}(u,v).
\end{align}
Here $B_{i}(u,v)=B_{i,i-1}(u,v)+B_{i+1,i}(u,v)$ is a blending function
assigned to the $i$-th side; everything else is as before.

There are, however, more constraints on the parameters. For a point
on the $i$-th side:
\begin{align}
d_{i-1} & =s_{i}, & d_{i+1} & =1-s_{i},\label{eq:gc-param-constr}\\
d_{i-1}' & =s_{i}', & d_{i+1}' & =-s_{i}',\label{eq:gc-param-constr-2}
\end{align}
where the prime symbol means all directional derivatives. These are
satisfied by a blended construction\textemdash details can be found
in the original paper~\cite{Salvi:2014}.

Because of this constrained parameterization, there are less ribbon
evaluations, and thus the computational cost of this surface is about
25\% less than that of the Gregory patch, while there is no noticeable
change in the surface~\cite{Salvi:2015:KEPAF}.

\subsection{\label{subsec:Midpoint-patch}Midpoint patch}

Here a new degree of freedom was added to the Gregory patch, in form
of a central control point $P_{0}$:
\begin{equation}
S_{MP}(u,v)=\sum_{i=1}^{n}I_{i,i-1}(u,v)B_{i,i-1}^{*}(u,v)+P_{0}B_{0}^{*}(u,v).
\end{equation}
The modified blending functions are defined as
\begin{align}
B_{i,i-1}^{*}(u,v) & =\frac{d_{i}H(1-s_{i-1})H(d_{i-1})+d_{i-1}H(s_{i})H(d_{i})}{d_{i}+d_{i-1}},\nonumber \\
B_{0}^{*}(u,v) & =1-\sum_{i=1}^{n}B_{i,i-1}^{*}(u,v),
\end{align}
where $H(x)$ is a Hermite blend
\begin{equation}
H(x)=(1-x)^{3}+3(1-x)^{2}x.
\end{equation}
Note that by definition $\sum_{i=1}^{n}B_{i,i-1}^{*}(u,v)+B_{0}^{*}(u,v)=1$.

The control point $P_{0}$ has the default location
\begin{equation}
P_{0}=\frac{1}{n}\sum_{i=1}^{n}I_{i,i-1}(0.5,0.5),
\end{equation}
but it can be used to move the center of the surface.

In this scheme we also have a new constraint on the parameterization.
For points on side $i$, it should satisfy
\begin{equation}
d_{j}=1.\quad j\notin\{i-1,i,i+1\}\label{eq:mp-param-constr}
\end{equation}
This can be achieved using generalized barycentric coordinates. Let
us look at this in detail, as this will be the base of our new parameterization
in Section~\ref{subsec:Constrained-barycentric}.

\subsection{\label{subsec:Barycentric-parameterization}Barycentric parameterization}

\begin{figure}
  \centering{
\subfloat[\label{fig:Barycentric-parameterization}Barycentric parameterization]{\begin{centering}
\includegraphics[width=0.46\textwidth]{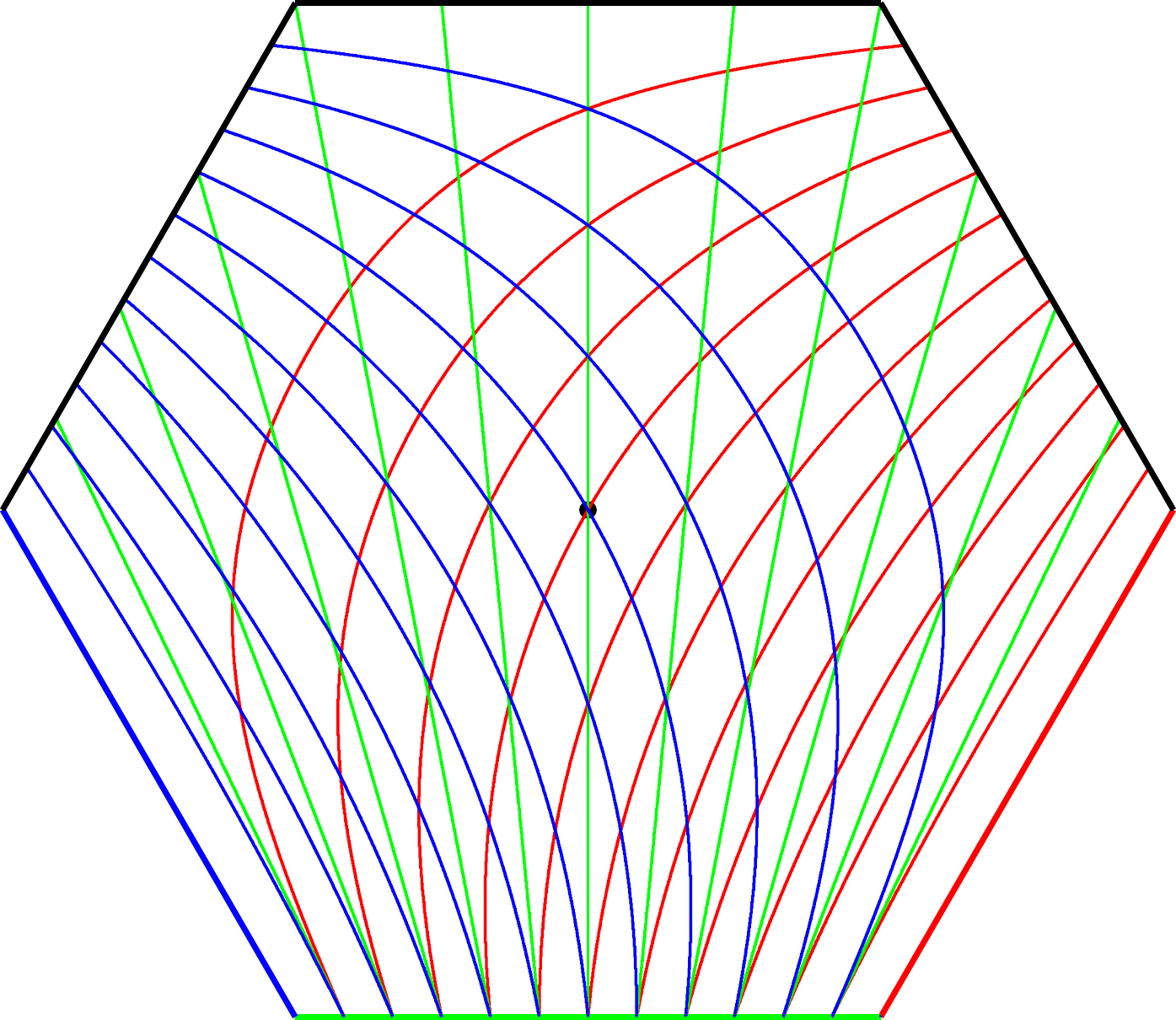}
\par\end{centering}
}\hfill
\subfloat[\label{fig:Constrained-barycentric}Constrained barycentric parameterization]{\begin{centering}
\includegraphics[width=0.46\textwidth]{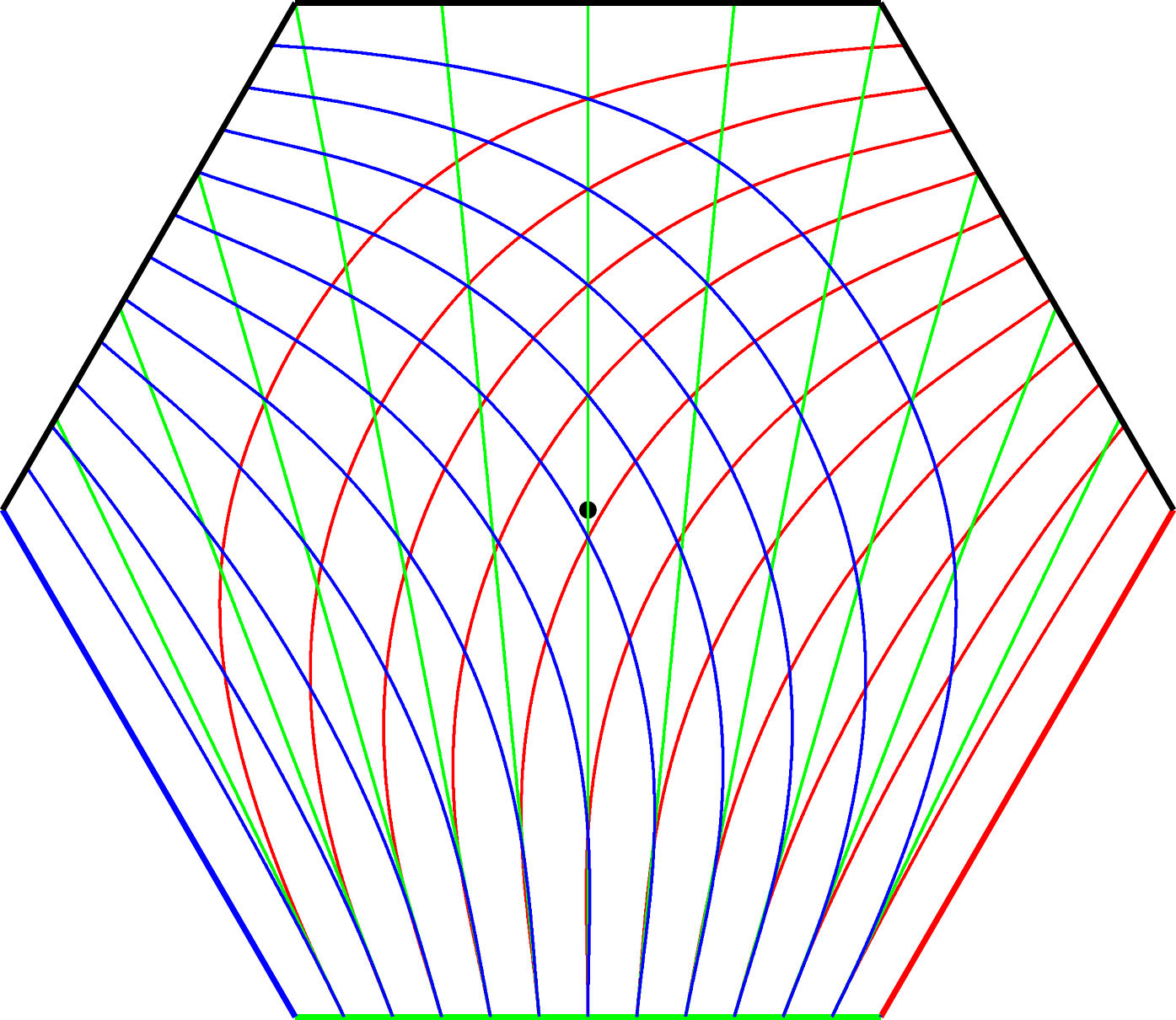}
\par\end{centering}
  }}
\caption{Constant parameter lines in a 6-sided domain. Side parameters of the
bottom side (green), and distance parameters of the left and right
sides (blue and red) are shown.}

\end{figure}

Given a convex polygon with vertices $V_{i}$, the Wachspress coordinates
$\{\lambda_{i}\}$ of a point $(u,v)$ have the following properties:
\begin{align}
\sum_{i=1}^{n}\lambda_{i}(u,v) & =1, & \lambda_{i}(V_{j}) & =\delta_{ij},
\end{align}
and $\lambda_{i}$ decreases linearly on the adjacent domain edges.
Now we can define side and distance parameters as
\begin{align}
s_{i} & =\lambda_{i}/\left(\lambda_{i-1}+\lambda_{i}\right), & d_{i} & =1-\lambda_{i-1}-\lambda_{i}.
\end{align}
It is easy to see that this construction satisfies all requirements.
An example is shown in Figure~\ref{fig:Barycentric-parameterization}.

\section{\label{sec:Midpoint-Coons-patches}Midpoint Coons patches}

We would like to combine the GC and MP patches, so that we have a
computationally efficient transfinite patch with fullness control.
The idea is simple: use the modified blending functions $B_{i,i-1}^{*}$
in the GC scheme:
\begin{align}
S_{MC}(u,v) & =\sum_{i=1}^{n}R_{i}(s_{i},d_{i})\left[B_{i,i-1}^{*}(u,v)+B_{i+1,i}^{*}(u,v)\right]\nonumber \\
 & -\sum_{i=1}^{n}Q_{i,i-1}(s_{i},s_{i-1})B_{i,i-1}^{*}(u,v).
\end{align}

Unfortunately this is not enough. The problem is in the parameterization:
it has to satisfy both Eqs.~(\ref{eq:gc-param-constr})\textendash (\ref{eq:gc-param-constr-2})
and~(\ref{eq:mp-param-constr}), which none of the original formulations
were capable of.

\subsection{\label{subsec:Constrained-barycentric}Constrained barycentric parameterization}

The barycentric parameters described in Section~\ref{subsec:Barycentric-parameterization}
already satisfy every constraint \emph{except} Eq.~(\ref{eq:gc-param-constr-2}).
This is evident in Figure~\ref{fig:Barycentric-parameterization},
as the blue and red constant parameter lines do not have the same
tangent as the green ones near the bottom.

Looking at this figure, we can see that what we need is a new distance
parameter $\hat{d}_{i}$ that ``behaves'' as $s_{i-1}$ when $s_{i}=0$,
and as $s_{i+1}$ when $s_{i}=1$; but we want to retain the original
$d_{i}$ for $d_{i}=0$ and $d_{i}=1$ (i.e., at the base side and
at non-adjacent sides).

Note that this is a very similar problem to transfinite surfaces\textemdash we
have four one-dimensional boundary constraints to be interpolated,
while for points inside the domain blended values are needed. Indeed,
it can be solved using the blending functions of Kato's patch~\cite{Kato:1991}.

Let the values at the boundaries be
\begin{align}
x_{1} & =d_{i}, & x_{2} & =s_{i+1}, & x_{3} & =d, & x_{4} & =1-s_{i-1},
\end{align}
and the parameters in the corresponding square domain be
\begin{align}
t_{1} & =d_{i}, & t_{2} & =1-s_{i}, & t_{3} & =1-d_{i}, & t_{4} & =s_{i}.
\end{align}
Then the new distance parameter is defined as
\begin{equation}
\hat{d}_{i}=\sum_{j=1}^{4}x_{j}\hat{B}_{j}(u,v),
\end{equation}
where
\begin{equation}
\hat{B}_{j}(u,v)=\frac{\prod_{l\neq j}t_{l}^{2}}{\sum_{k=1}^{4}\prod_{l\neq k}t_{l}^{2}}=\frac{1/t_{j}^{2}}{\sum_{k=1}^{4}1/t_{k}^{2}}.
\end{equation}
Note that this function is singular when $t_{i}=t_{j}=0$, $i\neq j$.
This does not present a problem, as the parameterization is well-defined
in these points.

This $\hat{d}_{i}$, combined with the original side parameter $s_{i}$,
gives a parameterization that satisfies all constraints. Figure~\ref{fig:Constrained-barycentric}
shows the parameterization in a 6-sided domain.

\section{\label{sec:Test-results}Test results}

\begin{table*}
\begin{centering}
\begin{tabular}{|c|c|c|c|c|c|c|}
\hline 
 & $n=3$ & $n=4$ & $n=5$ & $n=6$ & $n=7$ & $n=8$\tabularnewline
\hline 
\hline 
CB & 429ms & 316ms & 652ms & 760ms & 887ms & 968ms\tabularnewline
\hline 
GC & 321ms & 276ms & 466ms & 536ms & 616ms & 673ms\tabularnewline
\hline 
MP & 419ms & 341ms & 638ms & 752ms & 868ms & 953ms\tabularnewline
\hline 
MC & 299ms & 277ms & 441ms & 518ms & 578ms & 636ms\tabularnewline
\hline 
Speedup & 28.6\% & 18.8\% & 30.9\% & 31.1\% & 33.4\% & 33.3\%\tabularnewline
\hline 
\end{tabular}
\par\end{centering}
\caption{\label{tab:Evaluation-speed}Evaluation speed of different surface
representations on a 2.8GHz machine, with a resolution of ca.~10000
triangles, showing the speedup between MP and MC patches.}
\end{table*}

\begin{figure}[b]
\begin{centering}
\includegraphics[width=0.4\textwidth]{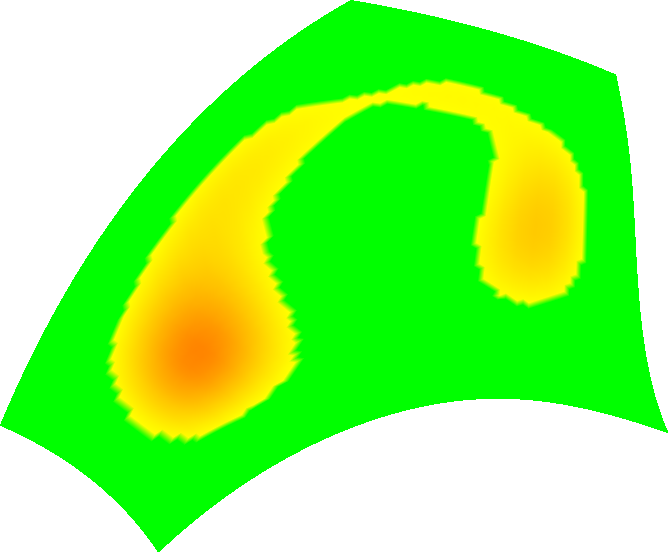}
\par\end{centering}
\caption{\label{fig:Deviation-between}Deviations between 5-sided MP and MC
patches. Full red color means $0.5\%$ of the bounding box axis. The
green region has deviation less than $0.2\%$.}
\end{figure}
As expected, the new MC patch shows an average of 30\% speedup compared
to the MP patch, see Table~\ref{tab:Evaluation-speed}.

Figure~\ref{fig:Deviation-between} shows the deviation between an
MP and an MC patch. The maximum deviation is $\approx0.4\%$ of the
bounding box axis, with an average deviation of $\approx0.1\%$. (These
values are even tighter, if the MP patch uses the same constrained
parameterization.) The two patches are visually indistinguishable.

\begin{figure}
\begin{centering}
\subfloat[Mean curvature map]{\begin{centering}
\noindent\begin{minipage}[t]{1\columnwidth}%
\includegraphics[width=0.45\textwidth]{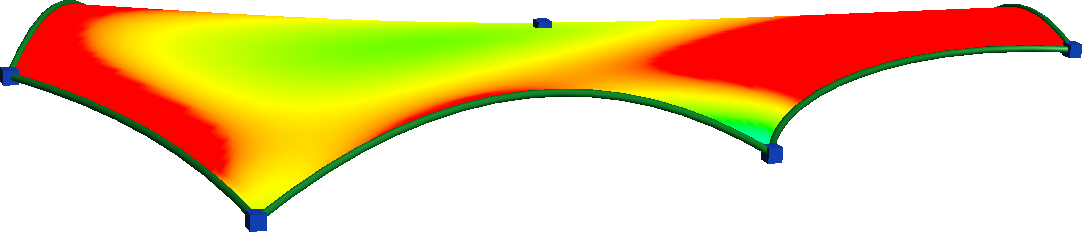}\hfill
\includegraphics[width=0.45\textwidth]{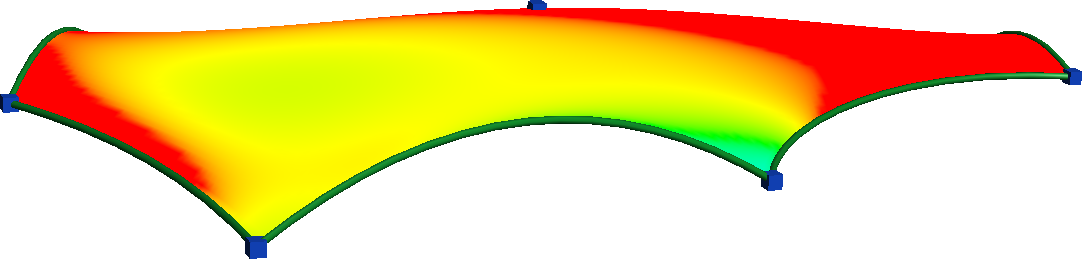}
\end{minipage}
\par\end{centering}

}
\par\end{centering}
\begin{centering}
\subfloat[Contouring]{\begin{centering}
\noindent\begin{minipage}[t]{1\columnwidth}%
\includegraphics[width=0.4\textwidth]{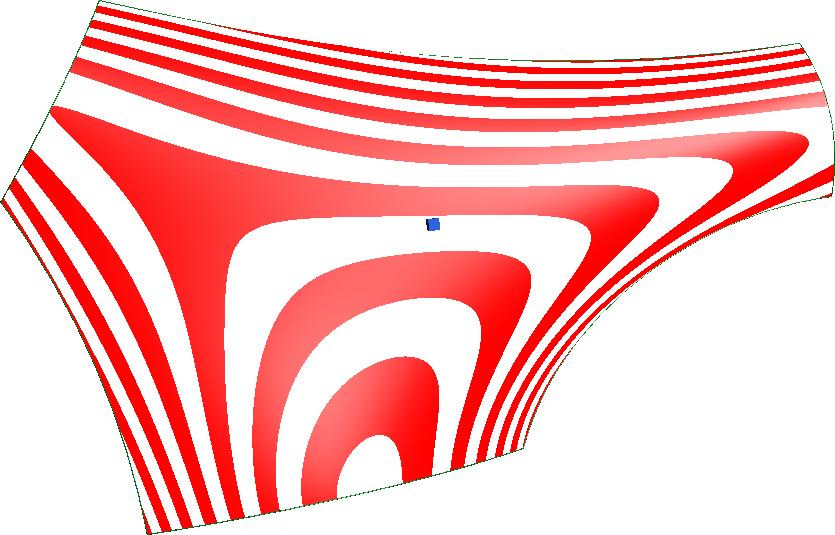}\hfill
\includegraphics[width=0.4\textwidth]{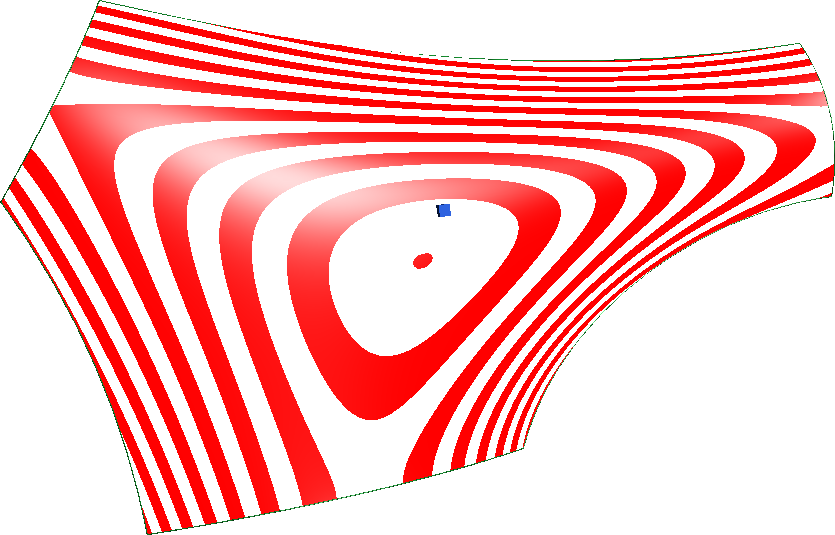}
\end{minipage}
\par\end{centering}

}
\par\end{centering}
\caption{\label{fig:Changing-the-fullness}Fullness change. The central control point is shown by a small cube.}

\end{figure}

In Figure~\ref{fig:Changing-the-fullness} we can see how the central
control point affects the shape of the surface. This change has no
effect on the $G^{1}$ interpolation properties, as can be seen from
the contourings.

\section*{Conclusion}

We have successfully combined two transfinite surface representations
into a new one that takes the best of both worlds: fast evaluation
and the ability to control the surface interior. Future work includes
optimization with the GPU and the development of efficient derivative
computation.

\bibliographystyle{plain}
\bibliography{cikkek,sajat}

\end{document}